# FEDERATED IDENTITY MANAGEMENT (FIdM) SYSTEMS LIMITATION AND SOLUTIONS

Maha Aldosary and Norah Alqahtani

Department of Computer Sciences, Imam Mohammad Ibn
Saud Islamic University, Riyadh, KSA

## ABSTRACT

*Efficient identity management system has become one of the fundamental requirements for ensuring safe, secure, and transparent use of identifiable information and attributes. FIdM allows users to distribute their identity information across security domains which increase the portability of their digital identities. However, it also raises new architectural challenges and significant security and privacy issues that need to be mitigated. In this paper, we presented the limitations and risks in Federated Identity Management system and discuss the results and proposed solutions.*

## 1. INTRODUCTION

Federated Identity Management (FIdM) is a concept that helps to link user's digital identities and attributes stored in several sits also allows cooperation on identity processes, policies, and technologies among various domains to simplifies the user experience. FIdM typically involves Identity Providers (IdPs) and Service Providers (SPs) in a trust structure called Circle of Trust (CoT) based on a business agreement where all the identifiable information of users are federated at a central location such as the Identity Provider IdPs who is responsible to pass authentication tokens to SPs, and SPs after that provide their resource to the user. FIdM is considered a promising approach to facilitate secure resource sharing among collaborating participants in heterogeneous IT environments [1].

 Many advantages demonstrated by Federated Identity Management systems such as reduce the cost provide convenience for the users and interoperability among Identity Management systems in addition to support single sign-on SSO service and other valuable services. However, it has limitations that provide several real security and privacy risks Due to the valuable information shared across domains in The FIdM using loosely coupled network protocols. The risks and limitations in FIdM require to be introduced and explained to find Appropriate solutions to mitigate these risks.

In this paper, we discussed the concept of personal identity in a real-world and digital identity as a prelude to the identity management systems. The notion of Identity Federation was discussed in this work as well some Federated Identity Management Architectures such as Liberty Alliance,

Security Assertion Mark-up Language SAML V2.0, WS-federation, and Shibboleth, etc. In this paper, we presented the limitations of Federated Identity Management based on how it affects the user. Finally, we discussed the solutions proposed to mitigate the risk of these limitations.

This paper is organised as follows: Section 2 gives background and basic information that needs to be understood before discussing the FIdM system. The concept of identity federation and the number of architectures that implement FIdM is given in section 3. Section 4 presented the limitation and risks in the FIdM. In section 5 provides a discussion of the solutions before the paper is concluded in section 6.

## 2. BACKGROUND

### 2.1. Identity

Human identity is a representation of an individual by several properties which indicates that person, reflecting its uniqueness, and distinguish that person from others. These properties could be intrinsic (e.g. DNA, retina scan, fingerprint), descriptive (e.g. name, birthplace, birthdate), demographic (e.g. gender, occupation), geographic (e.g. country, address, postcode) or psychographics (e.g. preferences, interests).[2]

The identity of an individual consists of a large number of personal properties. All subsets of the properties form partial identities of the person.[3] The person may have multiple different partial identities depending on the context. These partial identities could relate to roles the person plays. Identity involves all the primary characteristics that make each person unique but also all the characteristics that enable belonging to a particular group as well as established position within the group [4].

In today's world, living and working in the networked environment requires digital identity for each individual, it has allowed us to interact, transact, communicate, share reputations, and create trusted relationships with devices, people, and business electronically. Digital identity is the representation of identity in a digital system, Roussos et al [4] describe the digital identity as the electronic representation of personal information of an individual or organization (name, phone numbers, address, demographics, etc.).

Despite that there is a strong association between real life and digital identity, digital identity breaks from the restriction of everyday life, allowing users to exceed the boundaries of the real world[5]. clarify that digital environments granted the users the chance to get rid of the human qualities of age, race, gender, and disability.

### 2.2. Identity Management

Identity management (IdM) is defined as a set of procedures, policies and technologies that help authoritative sources as well as individual entities to manage and use identity information, it also provides access and privileges to end-users through authentication schemes [6]. Identity management procedures include management of the identity lifecycle, management of identity information, and management of entity authentication as an initial step for authorization.

Identity Management responsible for handling the lifecycle of identity, its creation, maintenance and eliminating a digital identity, by providing the credentials and means for identification during the preparatory process, through to authenticating and authorising access to resources, and to revoking access credentials and identities. Identity management is a crucial part of many security services since it assures user legitimacy. Therefore, identity management is an integral part of any access management system [7].

There are numerous technologies, services and terms related to identity management such as Directory services, Service Providers, Identity Providers, Digital Cards, Digital Identities, Web Services, Access control, Password Managers, Single Sign-on, Security Token Services, Security Tokens, WS-Trust, WS-Security, OpenID, OAuth, SAML 2.0 and RBAC.

Identity management is particularly used to authenticate a user on a system and make certain whether that user is allowed or unauthorised to access a particular system. IdM also covers issues such as how users obtain an identity, the protection of that identity and the technologies supporting that protection. Digital identity management technology is an essential function in enhancing and customizing the network user experience, protecting privacy, underpinning accountability in transactions and interactions, and respecting regulatory controls [8].

## 3. IDENTITY FEDERATION

Federated identity management (FIdM) is when multiple enterprises allow individuals to use the same identification information or login credentials to obtain access to the services or networks of all the enterprises in the group. The partners in a FIdM system are accountable for authenticating their users and for insuring for their access to the networks.

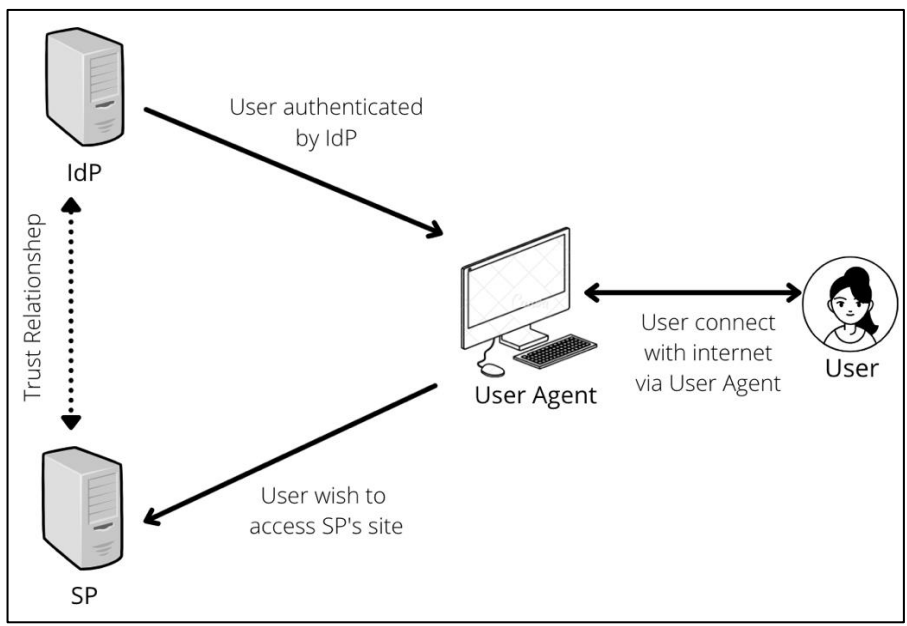

Figure 1. Component of FIdM system

The federated identity model includes four logical components: [9]

- A user is a person who acquires a specific digital identity to interact with an online network application.
- The user agent is a software application or browser that runs on any device such as PC, mobile phone and medical device. The online interactions of a user always take place via an agent, which can allow identity information flow or mediate it.

- The service provider (SP) site is a Web application that offloads authentication to a third party, which also might send the SP some user attributes. Because the SP depends on external information, it's often called a relying party (RP).
- The identity provider (IdP) is a Web site that users log in to and that occasionally stores attributes of common interest to share with several SPs.

In a Federated identity management system, the user might have one or more local identities issued by service providers (SPs), in addition to a single identity issued by the identity provider (IdP) within a specific domain called a circle of trust (CoT). A standard CoT composed of a single IdP and multiple SPs. In CoT, the IdPs must be trusted by all the SPs within it. Each SP could be a member of more than one CoT. A user can federate its IdP-issued identity with the local identities issued by SPs within the same CoT. [7]

With FIdM the user's credentials are always stored by the IdP. When a user registers into service, they do not have to provide their credentials to any of the SPs. Instead of authenticating directly with the user, the SP trusts the IdP to verify the user's credentials. The IdP then authorises the user to the application of SP, and the user is then allowed to access the service. Therefore, the user in FIdM systems never provides their credentials to anyone but the IdP.

FIdM presents numerous benefits to the various stakeholders, it offers users the single sign-on (SSO) capability that allows them to proceed between the various SPs with no need to authenticate or login again, it allows SPs to offload the cost of managing user attributes, passwords and login credentials to trusted IdPs, it provides scalability, allowing SPs to provide services to a greater number of users, it allows IdPs to maintain close relationships with end-users and sell them more services, as well as extract fees from the SPs they support [10].

Table 1. A Comparison between FIdM and SSO

|  | **FIdM** | **SSO** |
|---|---|---|
| **Single access** | To multiple system across various organization | To different services within a single organization |
| **User credential** | Given only to Idp | Given to any system the user logging into |
| **Log-in to several services** | Allow | Allow |
| **Use of the same credential** | Allow | Allow |
| **Authentication process** | Only once in the same working session | Only once in the same working session |
| **Identity federation** | Supported | Not supported |

FIdM has aspects that are similar to single sign-on (SSO), but they are different at their core. FIdM gives you SSO, but SSO does not necessarily give you FIdM. Despite that the SSO and FIdM both allow users to log in to several services using the same login credentials, there are two things that FIdM does that SSO cannot: Firstly, SSO allows users to access multiple systems only within a single organization, whereas FIdM allows users to log into systems across various organizations. Secondly, FIdM is more secure than SSO. For SSO, the user credentials are still being provided to any system that the user is logging into. While with FIdM, user's credentials are only given to the IdP exclusively.

Certainly, FIdM depends heavily on SSO technologies to authenticate the users across diverse websites and apps, however it has advanced these technologies further. Therefore, while FIdM does provide users SSO, SSO does not offer all of the benefits that FIdM does. Table 1 present a comparison between FIdM and SSO.

Federated identity management presents economic and convenience advantages to both the users and the organizations that employ it. However, there are some serious security considerations, techniques like strong authentication must be implemented for a secure SSO because SSO system may introduce a single-point-of-failure. [9] Moreover, FIdM requires a lot of trust and open communication between partners that choose to make use of it. Organizations that are considering creating or joining an identity federation need to assure that they agree upon all factors. [10]

### 3.1. Federated Identity Management Architectures

### 3.1.1. Liberty alliance

Liberty Alliance is a project presented first in 2001. according to the official web site of the project [11], it is a consortium of more than 150 member includes governments and companies from around the world. The consortium is committed to creating an infrastructure that provides support for all existing and emerging network access devices and has defined interoperability requirements developing an open standard for federated network identity for products that meet its specifications. The specifications developed by the Liberty Alliance Project enable individuals and organizations to control their identity information securely also it is providing conveniently by supporting single sign-on (SSO) service which is the service that enables users to interact with different service providers or Web sites with trust relationships by signing in just once.

The main objectives of the Liberty Alliance Project Specifications are to Serve as open standards for SSO, management of federated identity, and web services. Also, it aims to promote permission-based sharing of personal identity attributes and Enable consumers to protect their network identity information. Additionally, aims to create an open network identity infrastructure that supports all current and emerging user agents.

The specifications in the Liberty Alliance are enclouding the following components: Liberty Identity Federation Framework, Liberty Identity Web Services Framework, Liberty Identity Service Interface Specifications, Schema Files and Service Definition Documents, and Support Documents. They are developed to enable federated network identity management. Using web redirection and open-source technologies such as SOAP and XML, they enable distributed, cross-domain interactions [12][13].

For more information about Liberty Alliance: [14] [15] [16] [17]

### 3.1.2. Shibboleth

Shibboleth is a project created first by US-based Internet2 in 2003. It developed an open-source, standards-based system that provides access management for individuals to a resource depending on their role instead of their identities which means that Role-based attributes are used in the Shibboleth system. Shibboleth allows the affiliated institution of the user to authenticate the user to permit access to on-campus applications and the resources licensed by the library from service providers [18]. to protect the user's privacy, Shibboleth sends anonymous identification to the service provider.

Additionally, Shibboleth provides authorisation service that helps sites to make decisions for individual's access and privileges in online resources by transport the role attributes securely between the Identity Provider site (affiliated institution) and Resource Provider site to determine whether the user has a right to access the resource or not. Single sign-on (SSO) feature is supported in the Shibboleth system which makes the system more flexible and convenient.

For more information about Liberty Alliance: [14] [19] [20] [21]

### 3.1.3. WS-Federation

The Web Services Security Framework is an identity management approach proposed by International Business Machines Corporation and Microsoft Corporation with other companies. As in Liberty Alliance, they provide several specifications such as WS-Security, WS-Trust, and WS-Security Policy. These specifications determine how to control the assertions (security tokens) that contain identifiable information about the user and issued by an identity provider. The security tokens help the service provider SP to decide to wither or not the user have a right to access the service resource.

According to [22] [23], WS-Federation builds upon the base WSS specifications to define mechanisms which enable resources to be shared securely between different domains. The specifications introduce many services include security token service STS which is the IdP service that issues identity tokens to users based on their authentication. Authorisation service which decides giving access right to the user.

For more information about WS-Federation: [24] [25]

### 3.1.4. Security Assertion Mark-up Language SAML V2.0.

The first release of the Security Assertion Mark-up Language SAML was in 2002 by the Organization for the Advancement of Structured Information Standards OASIS. SAML is standard based on Extensible Mark-up Language XML helps to manage the authentication and authorisation processes between identity providers and service providers. The SAML system includes four main concepts which are assertions, protocols, profiles and bindings. Where assertion is the declaration user information asserted by the identity provider IdP for a service provider SP. The SAML protocol is helping to determine the rules on how to embed the SAML elements inside the request/response packet and on how to process them.

Transporting protocol messages using existing widely deployed communication protocols like HTTP (Hypertext Transfer Protocol) or SOAP (Simple Object Access Protocol) was described by The SAML bindings specification (SAMLBind). besides, The SAML profile specification (SAMLProf) provides many profiles that describe how the SAML elements can be used to implement a use case and achieve interoperability. [26] [27] [28]

### 3.1.5. Other Architectures

In addition to the FIdM Architectures that we talked about above, there are other federated architectures that designed originally for relatively simple applications such as OpenID [29] Which is open source user-centric and decentralized Identity management system. OpenID connect [30] Which is a simple identity layer on top of the OAuth 2.0 specifications family. It is the third generation of OpenID technology. It helps the SP to authenticate the End-User based on the authentication performed by an Authorisation Server, as well as to obtain user attribute in an interoperable and REST-like manner. Besides, SCIM (System for Cross-domain Identity

Management) Which is a specification designed for cloud-based applications to manage user identities and services [31].In Table 2, we provide a comparison between Liberty alliance, Shibboleth, Security Assertion Mark-up Language SAML V2.0., WS-Federation, OpenID and OpenID connect about the target area, storage of Identity information, Single Sign-On, Single Log-Out, Identity Mapping, Security Tokens and Access to web applications. [32] [33]

Table 2. Comparison Between FIdM Architectures

| | **Liberty alliance** | **SAML V2.0.** | **WS-Federation** | **Shibboleth** | **OpenID connect** | **OpenID** |
|---|---|---|---|---|---|---|
| **Identity mapping** | By opaque identifiers | Via pseudonym service | Via pseudonym service | By short-term random IDs | Using (STS) chains and JavaScript mapping | Using (STS) chains and JavaScript mapping |
| **Area targeted** | Business interactions | Business interactions | Business interactions | Digital academic resource sharing | Developer and programmer | Supporting developer and programmer |
| **Identity information storage** | User info could be distributed and federated | User info could be distributed and federated | User info could be distributed and federated | Centrally located and only attributes sent to SP | Attributes and info are distributed IdP | Attributes and info are distributed IdP |
| **Single sign-on** | Supported | Supported | Supported | Supported | Supported | Supported |
| **Single log-out** | Supported | Supported | Not supported | Not supported | Supported | Supported |
| **Security tokens** | Extends SAML assertions for Communicating authentication And authorisation security Tokens between providers | Extends SAML assertions for Communicating authentication And authorisation security Tokens between providers | Builds on ws-security's Profiles and Kerberos | Extend the IdP to support info card profiles using SAML assertions as security tokens | Use json security tokens (json web token) to communicate user attributes | Use json security tokens (json web token) for user attributes |

| Access to Services | Supports access of both Web Services and web applications | Supports access of both Web Services and web applications | Designed only for Web Services | Only supports access by web browsers to web apps | Support browser-based JavaScript, web app and native mobile apps | Support browser-based JavaScript, web app and native mobile apps |

## 4. LIMITATIONS AND DRAWBACKS IN FEDERATED IDENTITY MANAGEMENT SYSTEM

Federated Identity Management (FIdM) is a technique that allows the participating entities i.e., Service Providers (SPs) and Identity Providers (IdPs), to collaborate, on identity operations, technologies, and policies. FIdM also enables users of heterogeneous IT environments to share each other's resources. [34]. All the user identities in a FIdM system are federated at a central position, i.e., the Identity Provider (IdP). IdPs are responsible to proceed the authentication tokens to SPs, and after that SPs can provide their services to the requestor i.e., the user. It is also possible that the user has accounts with various IdPs, and the SP communicates with the relevant IdP for the set of attributes required. [35]. While FIdM is in general seen as a good thing, it does have some disadvantages. Based on how it effects user we determine the following limitations:

### 4.1. Trust

Any Federated Identity system is based fundamentally on mutual trust. The interactions in federated identity management systems occur only between pre-configured entities or closed circle of trust (CoT) due to the use of static establishment of trust which is the method where entities' trust relationship such as that between IdPs or SPs has to be pre-configured that done either during the registration phase to the system or via a trust negotiation process offline. Such limitation especially for a huge number of participating (IdPs and SPs) makes the system impractical, unscalable, and hard to establish trust relationship at runtime.[36] [37]

In any identity federation, each participating member must create and identify policies and security protocols which poses another challenge. Every member then is obligated to follow these rules, which may cause problems when various companies have different rules and requirements. Furthermore, since an organization can be a member of different federations, following these several policies and rules may become a challenge.

Current specifications of FIdM provides only the basic technical mechanisms to establish trust between participating members. However, they do not detail the requirements that need to be met before establishing these relationships.

### 4.2. Privacy

Privacy and data protection are a major concern in FIdM system due to personally identifiable information that shared between entities where the premier goal of FIdM models is to share identity attributes [34] there is no guarantee to prevent SPs and IdPs from misusing of identity information of users. Even though there are regulations such as [38] and [39] and privacy policies that protect the privacy of user's sensitive data but unfortunately there are no requirements to enforce these regulations and policies. Furthermore, many studies [35] [40][41] Proved that many SPs and IdPs sites are collecting, processing, and sharing data of users without user consent.

### 4.3. IdP discovery

The IdP discovery is the process of determining where authentication requests are going to be forwarded when a user wants to access an identity-based service. [40] One of the major significant security limitations in most of FIdM standards such as Shibboleth, Liberty and OpenID is that IdP discovery is performed on the SP server. This limitation could be exploited by a malicious SP to redirect a user to a web site masquerading as the IdP, which could then acquire the user's security credentials. [42]

Furthermore, FIdM systems rely on the constant communication between individual users and a centralized identity provider (IdP) for purpose of authenticating and grant authorisation. If the metadata used to authenticate a user to the IdP was compromised, through leaks, or any sort of attacks such as phishing attacks, an adversary would gain the same access to the federated identity provides to all other participating members.

### 4.4. Lack of Attribute-Aggregation Support

Another limitation of FIdM systems is that users can only choose one of their IdPs in any single working session with an SP, after that the IdP sends authentication and attribute assertion to the SP. Therefore, authorisation is restricted to a subset of the user's identity attributes. This isn't sufficient especially for Web-based services. There is a huge need for a mechanism that allows users to aggregate attributes from multiple IdPs in a single service session. This model could effectively help to protect the user's identifiers and prevents IdPs from exchanging data about users without their permission. However, each IdP still know that a federated user has several attributes at the other IdP. [43]

In Liberty, only one IdP can be queried in a single working session, and for any IdP in shibboleth, the authorisation framework only allows a single attribute authority (i.e. the Attribute Authority Service (AAS)) to be queried for user attributes. OpenID is also suffering from a lack of attribute-aggregation support. [42]

### 4.5. Complexity for the User

The usage of online services and transactions is growing every day, it is becoming necessary to grant the users and the service providers the tools they needed to make more transactions and expand the available services and the level of interaction and trust. [44] A drawback of FIdM based on SAML is the complexity of the protocol and resulting effort for configuration. Another limitation is the complexity for the user, especially because of the need from the user to choose their IdP at the Discovery Service (DS) and the users have to remember which federations they belong to, along with username and password. On the user side, the management of the identity is getting more complicated if the user uses multiple federations. [45]

### 4.6. Security

Identity theft is a serious concern in FIdM.[1] Security issues regarding a stolen identity will affect all federation partners, credentials (e.g. username and password pairs) must be protected in federated systems.

Common attacks are the impersonation attacks with stolen credentials. FIdM enabled systems to authenticate service requests by a security token attached to the request message. Therefore, impersonation attack can also be conducted by stealing user's security token which has been authenticated, this token can be used to access resources in the federated environment. [34]

An important property of FIdM is single-sign-on (SSO). However, a crucial challenge was addressed by Madsen et al. [46] they claim that federated SSO makes the job of attackers easier. That because after the attackers conduct a successful identity theft within a federation, they could compromise resources of all federated SPs, which leads to exposure of critical data.

Another important aspect is message security, Improper message security result in concerns for identity theft. Regarding identity management, techniques to protect message confidentiality and integrity are crucial to protect sensitive identity attribute and prevent modification of identity attributes. According to Maler and Reed [40]. systems are vulnerable if it does not provide security tokens to service request messages, through digital signatures, and check the message integrity before use.

OpenID does not support any proof-of-rightful-possession methods, while in shibboleth the use of proof-of-rightful-possession methods is optional. Therefore, an IdP might not provide a user with the means to prove rightful possession of security token to an SP. Such an approach increases the risk of an attacker using a stolen token to earn access to SP resources. [42]

### 4.7. Revocation

In FIdM, revocation means disabling identity data, often represented as identity attributes in security tokens, therefore they can't be used for identification and authorisation purposes anymore. Current FIdM systems lack practical and efficient revocation techniques, this may lead to security violations. Revocation is an important issue in credential-based systems [44].

## 5. DISCUSSION AND RESULTS

This section presents the existing solutions for the challenges and limitations that been discussed in the previous section.In table 3 we provide a summary of the solutions suggested to each limitation discussed in section 4:

Table 3. Solution suggested for each limitation

| LIMITATION | SOLUTION SUGGESTED | REFERENCES |
|---|---|---|
| **Trust** | Dynamic trust establishment | [47] [ 48] [35] |
| | Independent trust establishment mechanisms | [34] |
| | Ensure identity trust through SAML credential | [49] |
| | Trusted Computing Technologies | [50] |
| | Identity assurance | [51] [52] |
| **Privacy** | Pseudonyms | [40] [53] |
| | Undetectability | [54] |
| | Unlinkability | [54] [44] |
| | Decentralized identity | [54] |
| | Privacy by design | [55] |
| **IdP discovery** | List of IdPs | [56] |
| **Lack of attribute-aggregation support** | Supporting attribute-aggregation | [43] |
| **Complexity for the user** | User-centric approaches | [56] [35] [52] |
| | Smart contract | [57] |
| **Security** | Encryption | [34] |
| | Digital Signature | [58] |
| | User identity distribution | [1] |
| | Zero-knowledge proofs | [1] |
| | Channel security | [59] |
| | Authorisation policies | [53] |
| **Revocation** | Limit token lifetime | [34] |

In systems like cloud computing systems or Web services trust relationship needs to be processed on-demand and at runtime which cannot be done in static trust establishment. So, the dynamic establishment of the trust relationship between entities (IdPs or SPs) in FIdM systems with the help of factors like data on the SLA and reputation of the IdP/ SP could solve such issue. In [47] and [48] a FIdM systems with dynamic trust establishment was proposed.

In this paper [35.] the researchers identified a set of factors that are fundamental for developing a holistic FIdM framework or model. These factors are Trust Management, Trust Establishment, User Privacy, Consistent User Access Rights across CoTs, Continuous Trust Monitoring, and Adaptation to Environmental or Unanticipated Changes. Based on these factors, they also presented a comparative analysis that helps identifies challenges and areas of improvements in FIdM. Choosing a Trust Management and Trust Establishment scheme depends on the user requirement, however, user privacy and alignment of user access rights across different CoTs need to be handled with both Trust Management and Trust Establishment schemes.

In this paper, [49] presented a trusted federated identity management mechanism. This mechanism helps to ensure identity trust through SAML credential, to guarantee the trustworthiness of the federated identity management procedure.

Trusted Computing Technologies can help to solve authentication, privacy and trust concerns in federated identity management systems. Khattak et al. in [50] have presented three threats in federated systems: Identity theft, Misuse of Information gathered by malicious IdPs and SPs, and trust relationship issues due to no or weak trust among users, IdPs and SPs. A Trusted platform (TP) is presented that confirms the rules of the Trusted Computing Platform Alliance (TCPA) specification to counter these threats. The presented framework can help to secure user privacy; however, it doesn't help for situations that unidentified at requirements engineering time [35.].

For preserving privacy and protect user identities, pseudonyms are an important technique, especially when multiple web services cooperate to provide an aggregated offering that requires user-attribute sharing. [40]

If SPs are trusted to link authorisation requests to identities, Pseudonymous authorisation is implemented by Project Liberty, OpenID, Passport, and Client-Side Federation. [53] However, If SPs aren't trusted with links between authorisation requests and identities, then anonymous authorisation is employed. Anonymous authorisation implemented by eliminating all unique identifiers from messages or credentials that the service provider doesn't explicitly require. For example, Shibboleth supports anonymous authorisation, although users can choose to reveal a persistent identifier. Project Liberty lets a service provider request an anonymous, temporary, identifier for a user if the service provider elects to support anonymous authorisation. [54]

Undetectability and Unlinkability are privacy properties that help to preserve user privacy. Undetectability means users' ability to conceal actions from other parties. While Unlinkability concerns hiding correlations between combinations of actions and identities either permanently or temporarily, making it impossible to recognize two separate usages of the same credential [44]. Whether the linking between two identities was between action and identity, or between two actions, the level of trust that users grant to other parties determine the most appropriate design choice. In Project Liberty, the IdPs with established business relations create Circles of Trust (CoT). Within a CoT, a user can choose to federate two identities, in this case, the IdPs exchange information and the identities are linked [54]

In [54] the researchers identify crucial design choices essential to current identity management systems. They adopt a privacy-driven approach, which focuses on three privacy properties: Undetectability of authorisation requests which is concealing the user actions, Unlikability which is concealing correlations between combinations of actions and identities, and Confidentiality which means enabling users' control over dissemination of their attributes.

The most appropriate choice if IdPs can be trusted only with attributes that are specifically issued to them but not trusted with identity linking is a decentralized identity management system in which various, distinct IdPs each function separately using different protocols and not aware of each other. This architecture lets users select which IdPs to trust with which attributes, and spread critical attributes across distinct IdPs, thus ensuring unlikability of distinct identities. Most existing identity management systems, including Idemix, PRIME, Shibboleth, Higgins, CardSpace, OpenID, P-IMS, and U-Prove have adopted this approach.[54]

Though FIdM has mitigated the significant privacy flaws of the current situation by a number of techniques such as pseudonymous authentication and limited attribute release, however at the same time it also introduces new privacy issues, essentially by centralizing user data and making the track of user behavior easy and to link data of the same user together.

To mitigate these privacy risks the design process of FIdM systems needs to consider privacy requirements from the start. R. Hörbe and W. Hötzendorfer in [55] focus on privacy by design requirements for FIdM systems. They presented a catalogue of privacy-related architectural requirements, joining up legal, business and system architecture viewpoints. Furthermore, the demonstration of concrete FIdM models showing how the requirements can be implemented in practice.

A common solution to the problem of IdP discovery is to provide a list of IdPs to the user from which the user must select the proper IdPs. however, this is could be a problem especially when the list of possible IdPs gets extensive and the user, who usually ignorant about these issues, must conduct a choice. This is called the "where are you from" problem and is a significant concern regarding usability. Rieger [56] mentions this problem and adds that because the users can be part of numerous federations this will complicate the situation more.

Liberty solves this problem by using the Liberty-Enabled Client (LEC) profile. This profile requires the participation of Liberty-Enabled User Agent (LEUA) to handle the messages sent and received during the federation and authentication processes. [42]

It would enhance the practicality of FIdM if SPs could acquire user attributes from multiple independent attribute authority to be used in association with a particular IdP. Supporting attribute-aggregation will help to solve the limitation which users is limited to choose one of their IdPs in any single working session with an SP. [42]

The Liberty Alliance was the first group to address the problem of attribute aggregation through its model of identity federation. In this model, the first IdP to authenticate the user inquires if the user prefers to be introduced to other IdPs in the federation. Afterwards, when the user authenticates to another IdP, it invites the user to federate its second identity with that from the first IdP. If the user consents, the two IdPs each create a random alias for the user and exchange secretly. Thus, neither IDP have knowledge of the user's true login identifier at the other IdP, but each can refer to the same user through the random aliases, and thereby aggregate the attributes. [43]

One solution to mitigate privacy concerns and the complexity of the user is to empower users to control their identities. Increasing users control over their information is a good solution to avoid the misuse of information and data leakage. A user-centric identity management system is developed essentially from the perspective of end-users, it aims to make the user task of managing digital identities easy by providing them with more control over their identities. [42] User-centric approaches extend the users' privacy as the user can decide which private information to send to the Consumer (e.g. SP) as in [56]. There are many advantages of user-centric identity management such as higher usability and privacy for the users, simplification of the protocol and the configuration compared to SAML-based federated identity management and helps to create trust among cloud service providers in a federated environment. [35]

In [44] the proposed system which enabling controlled access to and selective sharing of critical user attributes in FIdM solutions by integrating authenticated dictionary (ADT) into FIdM, this can help to develop a user-centric and user-friendly attribute sharing system.

In this work, [57] they presented an identity management system that provides FIdM such that a user can authenticate and transfer attributes to a relying party (RP) without the involvement of a credential service provider (CSP). They accomplish this by leveraging a smart contract running on a blockchain5. Their approach can increase privacy and reducing costs.

Regarding identity management, techniques to protect message confidentiality and integrity are essential to prevent compromisation of sensitive identity attributes or modification of identity attributes. This can be achieved through mechanisms such as encryption.[ 34]

Message security is essential in FIdM to prevent attackers and intermediaries from manipulating the messages that are in transit. Improper message security rises concern such as identity theft, false authentication, and unauthorised use of resources. Liberty Alliance specifications advised XML Digital Signature and Encryption [58] for encrypting a complete or a part of the SOAP message to preserve the integrity and confidentiality of its contents.

Bhargav-Spantzel et al. [1] recommended two kinds of techniques to protect the misuse of identity information: The distribution of user identity information among various entities and use techniques such as zero-knowledge proofs to prevent identity theft within an IdP or SP. They recommend that single central IdP is a problem in Shibboleth. Moreover, their work is also highlighting that Liberty does not consider untrusted SP or IdP within the specifications.

The availability of information in FIdM models can be ensured by having a common protocol or mechanism for communicating authentication and other information between parties and securing communication channels and messages. Channel security can be accomplished using protocols like TLS1.0/SSL3.0 or other protocols with security characteristics that are equivalent to TLS or SSL. However, these protocols can only provide security at the transport level and not at the message level. For channel security, Liberty specifications highly recommend TLS/SSL with well-known cypher suites [58].

FIdM requires communicating parties to provide controlled access to information to authorised users. Authorisation goal is to deals with what information a user has access to or which operations a user can perform. A permission-based attribute sharing mechanism, which enables users to specify authorisation policies on the information that they want to share is recommended by Liberty specifications. [53]

A common way to mitigate revocation challenges is to limit the security token lifetime. By reducing the time-to-live to seconds or minutes the vulnerability window in cases of compromisation of the token will minimise. However, this may reduce the systems' usability as the user must reauthenticate to obtain a new valid security token. On the opposite side, when token expiration is set for a longer period user will benefit from the seamlessness, but the risk of identity theft and compromising information will increase. [34]

## 6. CONCLUSIONS

In our paper, we discussed the concept of identity federations well some federated identity management architectures such as liberty alliance, security assertion markup language SAML v2.0, WS-Federation, and Shibboleth with a comparison between these architectures. Furthermore, we presented the limitations of federated identity management based on how it effects the user. We determine the following limitations trust, privacy, IdP discovery, lack of attribute-aggregation support, complexity for the user, security, and revocation. Finally, we discussed the solutions that proposed to mitigate the risk of these limitations.

In future work, an in-depth analysis of privacy, security, and trust challenges in a federated environment will be conducted. Also, we will propose a FIdM system taking into consideration the limitations and solutions we found in this paper.

**AUTHORS**

**Maha Aldosary** is currently pursuing the M.Sc. degree in information security with Imam Muhammad ibn Saud Islamic University. She graduated with a bachelor's degree in computer science from University of Tabuk. Her research interests include blockchain technology, IoT, identity management and information security.

**Norah Alqahtani** is currently pursuing the M.Sc. degree in information security with Imam Muhammad ibn Saud Islamic University. She graduated with a bachelor's degree in computer science from Shagra University. Her research interests include Cloud Computing, blockchain technology, identity management and information security.